\documentclass{article}
\usepackage{amssymb}
\usepackage{amsmath}

\input{tcilatex}

\begin{document}

\ \ \ \ \ \ \ \ \ \ \ \ \ \ \ \ \ \ \ \ \ \ \ \ \ \ \ \ \ \ \ \ \ \ \ \ \ \
\ \ \ \ \ \ \ \ \ \ \ \ \ \ \ \ \ \ \ \ \ \ \ \ \ \ \ \ \ \ \ \ \ \ \ \ \ \
\ \ \ \ \ \ \ \ \ \ \ \ \ \ \ \ \ \ \ \ \ \ \ \ \ \ \ \ \ \ \ \ TIFR/TH/02-26

\vspace*{1cm}

\begin{center}
{\Large \textbf{Three-Neutrino Mass Matrices with Two Texture Zeros}} \\[1cm]
Bipin R. Desai$^{1}$, D.P. Roy$^{2}$ and Alexander R. Vaucher$^{1}$
\end{center}

\vspace*{1cm}

\noindent $^1$Physics Department, University of California, Riverside,
California 92521, USA

\noindent $^2$Department of Theoretical Physics, Tata Institute of
Fundamental Research, Homi Bhabha Road, Mumbai 400 005, INDIA

\vspace*{1cm}

\begin{center}
Abstract
\end{center}

\bigskip

Out of the fifteen $3\times 3$ neutrino mass matrices with two texture zeros
seven are compatible with the neutrino oscillation data. While 2 of them
correspond to hierarchical neutrino masses and 1 to an inverted hierarchy,
the remaining 4 correspond to degenerate masses. Moreover only the first 3
of the 7 mass matrices are compatible with the maximal mixing angle of
atmospheric neutrino and hence favoured by data. We give compact expressions
for mass matrices in terms of mass eigenvalues and study phenomenological
implications for the 7 cases. \ Similarity of the textures of the neutrino,
charged-lepton mass matrices with those of quark mass matrices is also
discussed.\newpage

\bigskip\ \ \ \ \ \ \ \ \ \ \ \ \ \ \ \ \ \ \ \ \ \ \ \ \ \ \ \ \ \ \ \ \ \
\ \ \ \ \ \ \ \textbf{I. Introduction}

There has been a long standing interest in the texture zeros of the $3\times
3$ quark mass matrices as a possible source of the observed hierarchies in
their masses and mixing angles. In particular the up and down quark mass
matrices with 2 or 3 texture zeros are known to successfully relate the
ratio of quark masses to the mixing angles [1]. In the context of neutrino
physics also there is a long history of invoking texture zeros for
describing the $4\times 4$ [2] as well as the standard $3\times 3$ neutrino
[3,4,5] mass matrices. Recently Frampton, Glashow and Marfatia [6] have
systematically compared the predictions of all the symmetric $3\times 3$
neutrino mass matrices with two or more independent texture zeros with the
neutrino mass and mixing parameters as derived from the oscillation data. A
symmetric $3\times 3$ mass matrix has in general 6 independent elements.
They find no neutrino mass matrix with three or more texture zeros, which is
compatible with the neutrino oscillation data. Moreover they find that 7 out
of the 15 independent neutrino mass matrices with two textutre zeros ($%
^{6}C_{2}=15$) are compatible with the oscillation data. The present work
carries this investigation one step further, giving explicit expressions for
the mass matrices in terms of the three mass eigenvalues. We also give
numerical estimates of these mass eigen values for the 7 cases, which are
compatible with the neutrino oscillation data.

As usual the neutrino mass matrices shall be written in their flavour basis,
which corresponds to the mass basis of the charged leptons. For simplicity
we shall consider real mass matrices. One can easily check that the main
results are unaffected by the introduction of CP violating phases. On the
other hand the constraints on these phases arising from the texture zeros
have been recently discussed in [7]. A $3\times 3$ real symmetric mass
matrix with two texture zeros has four independent parameters. Three of them
can be determined in terms of the three mass eigenvalues using the
invariance of trace and determinant, i.e. 
\begin{equation*}
\mathrm{Tr.}\ \mathcal{M}=m_{1}+m_{2}+m_{3},\ \mathrm{Det.}\ \mathcal{M}%
=m_{1}m_{2}m_{3},
\end{equation*}%
\begin{equation}
\ \mathrm{Det.}\ \mathcal{M}\times \mathrm{Tr.}\ \mathcal{M}%
^{-1}=m_{1}m_{2}+m_{2}m_{3}+m_{3}m_{1}.  \tag{1}
\end{equation}%
But we need one experimental input to determine the remaining parameter. We
shall assume the maximal mixing angle for the atmospheric neutrino
oscillation [8] to provide this input, i.e. 
\begin{equation}
t_{1}=\tan \theta _{23}=1.  \tag{2}
\end{equation}%
This represents by far the most robust result of the neutrino oscillation
experiments so far. The most favoured value of this mixing parameter has
remained 1 over the years, while the error bar has shrunk steadily.
Therefore we treat this as a manifestation of an underlying symmetry of the
neutrino mass matrix. It is conceivable of course that the physical value of
this parameter may not be exactly 1. Nonetheless it is fair to assume that
this point is smoothly connected to the symmetry limit. Therefore we expect
the phenomenologically favoured mass matrix to be compatible with the
maximal mixing angle of eq. (2). We shall see in section III below that 4
out of the 7 experimentally allowed mass matrices with two texture zeros are
incompatible with this requirement.

\bigskip

The relation (1) was successfully used in determining the structure of the
up- and down- quark mass matrices in terms of the mass eigenvalues for two
and three texture zeroes [9]. \ The supplemental relation in this case came
from the assumption that the three angles were very small (mass hierarchy)
so that the triangular matrix technique could be implemented [9],[10]. \ We
have briefly discussed this in section IV and compared their textures with
those of the neutrino, charged-lepton system.

\bigskip

\qquad \qquad \qquad \qquad \qquad\ \ \textbf{II. Experimental Constraints}

We shall impose the following experimental constraints on the neutrino
masses and mixing angles.

\begin{enumerate}
\item[i)] The atmospheric neutrino data implies 
\begin{equation}
\Delta _{a}=|m_{3}^{2}-m_{1,2}^{2}|=(1.7-4)10^{-3}\ \mathrm{eV}^{2},  \tag{3}
\end{equation}%
\begin{equation}
\tan ^{2}\theta _{23}=t_{1}^{2}=0.9-1,  \tag{4}
\end{equation}%
at 90\% CL [8].

\item[ii)] The solar neutrino data admits LMS (LOW) solution at 90 (99.7\%)
CL with [11] 
\begin{equation}
\Delta _{s}=|m_{2}^{2}-m_{1}^{2}|\simeq 6\times 10^{-5}\ (1\times 10^{-7})\ 
\mathrm{eV}^{2},  \tag{5}
\end{equation}%
\begin{equation}
t_{3}^{2}=\tan ^{2}\theta _{12}\simeq 0.4\ (0.6).  \tag{6}
\end{equation}

\item[iii)] The CHOOZ and Paoloverde atomic reactor experiments give the
90\% CL limit [12] 
\begin{equation}
s_{2}=\sin \theta _{13}<0.16.  \tag{7}
\end{equation}
\end{enumerate}

Thus we shall require the ratio of the solar and atmospheric neutrino mass
differences to satisfy the inequality 
\begin{equation}
\Delta _{s}/\Delta _{a}=|m_{2}^{2}-m_{1}^{2}|/|m_{3}^{2}-m_{1,2}^{2}|\text{ }%
\lesssim 2\times 10^{-2}.  \tag{8}
\end{equation}

\bigskip

\ \ \ \ \ \ \ \ \ \ \ \ \ \textbf{III. Neutrino Mass Matrices}

The mass matrix can be expressed in terms of the above masses and mixing
angles as

\begin{equation}
\mathcal{M}=U%
\begin{pmatrix}
m_{1} & 0 & 0 \\ 
0 & m_{2} & 0 \\ 
0 & 0 & m_{3}%
\end{pmatrix}%
U^{T},  \tag{9}
\end{equation}%
where 
\begin{equation}
U=%
\begin{pmatrix}
c_{2}c_{3} & c_{2}s_{3} & s_{2} \\ 
-c_{1}s_{3}-s_{1}s_{2}c_{3} & c_{1}c_{3}-s_{1}s_{2}s_{3} & s_{1}c_{2} \\ 
s_{1}s_{3}-c_{1}s_{2}c_{3} & -s_{1}c_{3}-c_{1}s_{2}s_{3} & c_{1}c_{2}%
\end{pmatrix}%
.  \tag{10}
\end{equation}%
Using the maximal mixing constraint (2) we get the following expressions for
the matrix elements of $\mathcal{M}$, which are valid upto the first order
terms in $s_{2}$. 
\begin{eqnarray}
\mathcal{M}_{11} &=&m_{1}c_{3}^{2}+m_{2}s_{3}^{2},  \notag \\[0.08in]
\mathcal{M}_{12} &=&-(m_{1}-m_{2})s_{3}c_{3}/\sqrt{2}+m_{3}s_{2}/\sqrt{2}%
-(m_{1}c_{3}^{2}+m_{2}s_{3}^{2})s_{2}/\sqrt{2},  \notag \\[0.08in]
\mathcal{M}_{13} &=&(m_{1}-m_{2})s_{3}c_{3}/\sqrt{2}+m_{3}s_{2}/\sqrt{2}%
-(m_{1}c_{3}^{2}+m_{2}s_{3}^{2})s_{2}/\sqrt{2},  \notag \\[0.08in]
\mathcal{M}_{22}
&=&(m_{1}s_{3}^{2}+m_{2}c_{3}^{2})/2+m_{3}/2+(m_{1}-m_{2})s_{2}s_{3}c_{3}, 
\notag \\[0.08in]
\mathcal{M}_{33}
&=&(m_{1}s_{3}^{2}+m_{2}c_{3}^{2})/2+m_{3}/2-(m_{1}-m_{2})s_{2}s_{3}c_{3}, 
\notag \\[0.08in]
\mathcal{M}_{23} &=&-(m_{1}s_{3}^{2}+m_{2}c_{3}^{2})/2+m_{3}/2.  \TCItag{11}
\end{eqnarray}%
From these expressions one can study the implications of setting different
pairs of matrix elements to zero, as we see below. \bigskip 

\noindent \textbf{(A)} \underline{\textbf{Hierarchical Solutions}} \medskip

\begin{enumerate}
\item[1)] $\mathcal{M}_{11}=0$, $\mathcal{M}_{12}=0$: They imply 
\begin{equation}
t_{3}^{2}=-m_{1}/m_{2},\ s_{2}=\sqrt{-m_{1}m_{2}}/m_{3}.  \tag{12}
\end{equation}%
Combining these with eqs. (6) and (7) we get $|m_{1}|<|m_{2}|\ll m_{3}$,
i.e. hierarchical masses. From eqs. (3) and (5) we get $|m_{3}|\simeq .05$, $%
|m_{2}|\simeq .009$, $|m_{1}|\simeq .004$ eV for the LMA solution; while the 
$m_{1}$ and $m_{2}$ values are each suppressed by a little over an order of
magnitude for the LOW solution. In the former case we get $s_{2}\simeq 0.12$%
, i.e. close to the CHOOZ limit (7); while it is over an order of magnitude
smaller in the latter case. Note that for this solution the $\nu _{e}$
Majorana mass, 
\begin{equation}
\mathcal{M}_{11}=\sum U_{ei}^{2}m_{i},  \tag{13}
\end{equation}%
is zero. Thus it predicts no $0\nu \beta \beta $ signal even at the level of
.001 eV, which corresponds to the highest level of sensitivity expected at
the proposed GENIUS experiment [13].

Finally substituting (12) in (11) gives the explicit form of the mass matrix
to first order in $s_{2}$, i.e. 
\begin{equation}
\mathcal{M}=%
\begin{pmatrix}
0 & 0 & \sqrt{-2m_{1}m_{2}} \\ 
0 & \frac{m_{1}+m_{2}+m_{3}}{2} & \frac{m_{3}-m_{1}-m_{2}}{2} \\ 
\sqrt{-2m_{1}m_{2}} & \frac{m_{3}-m_{1}-m_{2}}{2} & \frac{m_{1}+m_{2}+m_{3}}{%
2}%
\end{pmatrix}
\tag{14}
\end{equation}%
One can also derive this directly from eqs. (1) and (2). A special case of
this mass matrix corresponding to bimaximal mixing, $t_{3}=-m_{1}/m_{2}=1$,
has been obtained in a $U(1)$ gauge extension of the standard model,
corresponding to the gauge charge $L_{\mu }-L_{\tau }$ [14]. However we
could find no simple dynamical model for the general case of $m_{1}\neq
-m_{2}$. Some explorations in this direction can be found in refs. [5] and
[15].

\item[2)] $\mathcal{M}_{11}=0$, $\mathcal{M}_{13}=0$: The phenomenological
predictions for this case are identical to the previous one except for the
change of sign of $s_{2}$. The mass matrix is simply obtained from (14) by
interchanging the 12 and 13 elements.
\end{enumerate}

\bigskip

\noindent \textbf{(B)} \underline{\textbf{Degenerate Solutions}} \medskip

\begin{enumerate}
\item[1)] $\mathcal{M}_{22}=0$, $\mathcal{M}_{13}=0$: Substituting these in
eq. (11) imply 
\begin{equation}
m_{1}^{2}=m_{3}^{2}+4s_{2}m_{3}^{2}/t_{3},\
m_{2}^{2}=m_{3}^{2}-4s_{2}m_{3}^{2}t_{3},  \tag{15}
\end{equation}%
i.e. $m_{1}^{2},m_{3}^{2},m_{2}^{2}$ are nearly degenerate and occur in that
order. Thus $|m_{2}^{2}-m_{1}^{2}|>|m_{3}^{2}-m_{2}^{2}|$, in gross
contradiction with the observed inequality of the solar and atmospheric
neutrino masses (eq. 8).

As shown in [6], this mass matrix is compatible with the experimental
constraint of eq. (8) away from the maximal mixing points (2), i.e. for $%
t_{1}\neq 1$. In this case one gets 
\begin{eqnarray}
m_{1}^{2} &=&m_{3}^{2} [t_{1}^{4}+2s_{2}t_{1}^{2}(t_{1}+1/t_{1})/t_{3}] , 
\notag \\[0.08in]
m_{2}^{2} &=&m_{3}^{2} [t_{1}^{4}-2s_{2}t_{1}^{2}(t_{1}+1/t_{1})t_{3}] . 
\TCItag{16}
\end{eqnarray}
Compatibiity with the experimental constraint of eq. (8) can be achieved for 
\begin{equation}
s_{2}\sim 0.5\times 10^{-2}(1-t_{1}^{4})/(t_{3}+1/t_{3})\lesssim 4\times
10^{-4},  \tag{17}
\end{equation}%
corresponding to the $t_{1}^{2} \geq 0.9$ range of eq. (4). Thus the $s_{2}$
angle in this case is too small to be observed at the future long base line
experiments. On the other hand the $\nu _{e}$ Majorana mass 
\begin{equation}
\mathcal{M}_{11}\simeq t_{1}^{2}\sqrt{{\frac{\Delta _{a}}{1-t_{1}^{4}}}}\geq
0.10\ \mathrm{eV}.  \tag{18}
\end{equation}%
This is not too far below the present experimental upper limit of 0.2 eV
[16], and will surely be measurable at GENIUS [13]. Note that this
represents the common mass scale of degenerate neutrinos and it is
cosmologically significant. Nonetheless we consider the incompatibility of
this solution with the maximal mixing region $(t_{1}=1)$, favoured by the
atmospheric neutrino data, to be a serious drawback of this model. The
expressions for the matrix elements are rather long for $t_{1}\neq 1$.
Therefore we are not displaying the explicit form of the mass matrix.

\item[2)] $\mathcal{M}_{33}=0$, $\mathcal{M}_{12}=0$: Substituting these in
eq. (11) imply 
\begin{equation}
m_{1}^{2}=m_{3}^{2}-4s_{2}m_{3}^{2}/t_{3},\
m_{2}^{2}=m_{3}^{2}+4s_{2}m_{3}^{2}t_{3},  \tag{19}
\end{equation}%
i.e. nearly degenerate $m_{1}^{2},m_{3}^{2},m_{2}^{2}$ with $%
|m_{2}^{2}-m_{1}^{2}|>|m_{3}^{2}-m_{2}^{2}|$ as in the previous case. In
fact the magnitudes of $s_{2}$ and $t_{3}$ are the same as above. The
results for $t_{1}\neq 1$ are similar to the previous case with $t_{1}$
replaced by $-1/t_{1}$. Consequently the predicted $s_{2}$ and $\nu _{e}$
Majorana mass are very close to those of eqs. (17) and (18).

\item[3)] $\mathcal{M}_{22} = 0$, $\mathcal{M}_{12} = 0$: The results are
practically the same as in 2).

\item[4)] $\mathcal{M}_{33}=0$, $\mathcal{M}_{13}=0$: The results are
practically the same as in 1).
\end{enumerate}

\bigskip \qquad Matrices $B_{1}$ through $B_{4}$ in terms of mass
eigenvalues are listed in Appendix I.

\bigskip

\noindent \textbf{(C)} \underline{\textbf{Inverted Hierarchy}} \medskip

$\mathcal{M}_{22}=0$, $\mathcal{M}_{33}=0$: Substituting these in eq. (11)
gives 
\begin{equation}
s_{2}=0,\ t_{3}^{2}=-(m_{2}+m_{3})/(m_{1}+m_{3}).  \tag{20}
\end{equation}%
The first equality implies of no observable CP violation in the neutrino
sector. The consistency of the second with eqs. (6) and (8) implies an
inverted hierarchy of masses, i.e. 
\begin{equation}
-m_{2}\simeq m_{1}\simeq (2-4)m_{3}.  \tag{21}
\end{equation}%
Substituting this in eq. (3) gives $-m_{2}\simeq m_{1}\simeq .05$ eV and $%
m_{3}\simeq .02$ eV. The predicted $\nu _{e}$ Majorana mass is 
\begin{equation}
\mathcal{M}_{11}\simeq {\frac{1-t_{3}^{2}}{2t_{3}}}\sqrt{\Delta _{a}}\simeq
(.025-.013)\ \mathrm{eV}.  \tag{22}
\end{equation}%
This is an order of magnitude below the present upper limit of 0.2 eV [15],
but will be measurable at GENIUS [13]. Substituting (20) in (11) gives the
explicit form of the mass matrix 
\begin{equation}
\mathcal{M}=%
\begin{pmatrix}
m_{1}+m_{2}+m_{3} & -\sqrt{-(m_{1}+m_{3})(m_{2}+m_{3})/2} & \sqrt{%
-(m_{1}+m_{3})(m_{2}+m_{3})/2} \\ 
-\sqrt{-(m_{1}+m_{3})(m_{2}+m_{3})/2} & 0 & m_{3} \\ 
\sqrt{-(m_{1}+m_{3})(m_{2}+m_{3})/2} & m_{3} & 0%
\end{pmatrix}%
.  \tag{23}
\end{equation}%
\bigskip

\noindent \textbf{(D)} \underline{\textbf{Disallowed Cases}} \medskip

Finally let us briefly indicate why the remaining 8 mass matrices are
experimentally disallowed.

\begin{enumerate}
\item[1)] $\mathcal{M}_{12}=0$, $\mathcal{M}_{13}=0$: $\mathcal{M}_{12}-%
\mathcal{M}_{13}=0$ implies 
\begin{equation}
(m_{1}-m_{2})\sin 2\theta _{12}=0,  \tag{24}
\end{equation}%
i.e. no solar neutrino oscillation.

\item[2)] $\mathcal{M}_{12}$ or $\mathcal{M}_{13}=0$, $\mathcal{M}_{23}=0$: $%
\sqrt{2}\mathcal{M}_{12(13)}-2s_{2}\mathcal{M}_{23}=0$ implies 
\begin{equation}
|\tan 2\theta _{12}|={\frac{2|t_{3}|}{1-t_{3}^{2}}}=2|s_{2}|<0.32,  \tag{25}
\end{equation}%
in gross disagreement with the solar neutrino result (6).

\item[3)] $\mathcal{M}_{11}=0$, $\mathcal{M}_{22}$ or $\mathcal{M}_{33}$ or $%
\mathcal{M}_{23}=0$: In each case one gets (neglecting $O(s_{2})$ terms) 
\begin{equation}
|m_{2}^{2}-m_{1}^{2}|=(1-t_{3}^{4})m_{2}^{2},\
|m_{3}^{2}-m_{2}^{2}|=(2t_{3}^{2}-t_{3}^{4})m_{2}^{2},  \tag{26}
\end{equation}%
in conflict with eqs. (6) and (8).

\item[4)] $\mathcal{M}_{22}$ or $\mathcal{M}_{33}=0$, $\mathcal{M}_{23}=0$: $%
\mathcal{M}_{22(33)} + \mathcal{M}_{23}=0$ implies

\begin{equation}
m_{3}=-(m_{1}-m_{2})s_{2}s_{3}c_{3},\ \mathrm{i.e.}\ m_{1}\simeq -m_{2}\gg
m_{3}  \tag{27}
\end{equation}%
using eqs. (7) and (8). Substituting this in $\mathcal{M}_{23}=0$ implies $%
t_{3}^{2}\simeq 1$, in conflict with eq. (6).

Matrices $D_{1}$ through $D_{8}$ in terms of mass eigenvalues are listed in
Appendix I.
\end{enumerate}

\bigskip

\textbf{\ \ \ \ \ \ \ \ \ \ \ \ \ \ IV. \ Comparison with Quark Mass Matrices%
}

\textbf{\bigskip }

\qquad As stated earlier, equation (1) was used in obtaining the structure
of the up- and down- quark mass matrices, U and D, with two and three
texture zeroes, in terms of the mass eigenvalues [9]. \ In determining their
matrix elements it was further assumed that the mixing angles were very
small so as to be compatible with mass hierarchy [9], [10]. \ Among the
available choices, the structures that fit the experiment the best, with two
texture zeroes, were [9]

\ \ \ \ \ \ \ \ \ \ \ \ \ \ \ \ \ \ \ \ \ \ \ \ \ \ \ \ \ \ \ 

\begin{equation}
U=%
\begin{pmatrix}
0 & 0 & \sqrt{-m_{1}m_{2}} \\ 
0 & m_{2} & \sqrt{m_{1}m_{3}} \\ 
\sqrt{-m_{1}m_{2}} & \sqrt{m_{1}m_{3}} & m_{3}%
\end{pmatrix}
\tag{28}
\end{equation}%
with $m_{1}=m_{u},~m_{2}=m_{c},~m_{3}=m_{t}$

and

\begin{equation}
D=%
\begin{pmatrix}
0 & \sqrt{-m_{1}m_{2}} & 0 \\ 
\sqrt{-m_{1}m_{2}} & m_{2} & \sqrt{m_{1}m_{3}} \\ 
0 & \sqrt{m_{1}m_{3}} & m_{3}%
\end{pmatrix}
\tag{29}
\end{equation}

\bigskip

with $m_{1}=m_{d},~m_{2}=m_{s},~m_{3}=m_{b}.$

These matrices together give the correct CKM angles [9]

\bigskip

\qquad It is interesting that the texture of the above $U$- matrix is
similar to the best choice for the neutrino mass matrix ($\mathcal{M}$) for $%
s_{2}>0$ , and hierarchical mass given by (14). \ One may conjecture that in
that case the texture of the charged-lepton mass matrix ($L$) should be the
same as the $D$-matrix. \ Indeed, since the hierarchy of masses in $L$ is
similar to $D$, the entire structure of the two matrices may be similar.

\bigskip

\qquad In other words, the best choice for the neutrino and charged-lepton
mass matrices would be (14) and

\bigskip

\begin{equation}
L=%
\begin{pmatrix}
0 & \sqrt{-m_{e}m_{\mu }} & 0 \\ 
\sqrt{-m_{e}m_{\mu }} & m_{\mu } & \sqrt{m_{e}m_{\tau }} \\ 
0 & \sqrt{m_{e}m_{\tau }} & m_{\tau }%
\end{pmatrix}
\tag{30}
\end{equation}

\bigskip

We realize that the neutrino mass matrices were written above in the basis
in which $L$ is diagonal whereas the above $L$ is not. \ One could
diagonalize $L$ by rotating it through appropriate angles. \ The new $%
\mathcal{M}$ matrix will then be different from matrix (14) by terms
proportional to the rotation angles which, however, are very small compared
to their neutrino counterparts since they are proportional to the ratios of
the charged lepton masses which are, indeed, very small. \ Even for $%
\mathcal{M}_{13}$, which is proportional to $s_{2}$ , and, therefore, small,
the new $\mathcal{M}_{13}$ will not be drastically different since the
relevent rotation angle here is proportional to $L_{13}$ which is actually
zero. \ Thus we expect the new $\mathcal{M}$ not to be very different from
(14), inspite of the fact that (30) is not diagonal.

\bigskip

\ \ \ \ \ \ \ \ \ \ \ \ \ \ \ \ \ \ \ \ \ \ \ \ \ \ \ \ \ \ \ \ \ \ \ \ \ 
\textbf{V. \ Conclusion}

\bigskip

Only a limited number of texture patterns for the neutrino mass matrices are
consistent with experiment, particularly if one takes the experimentally
favoured value of $t_{1}=1$. \ These are given by $A_{1},A_{2},$and $C$ . \
In particular, for hierarchical masses and $s_{2}>0$ there is only one
possible matrix $A_{1}$ given by (14). \ We also note that this particular
matrix has a texture which is similar to the up-quark matrix, that is found
compatible with experiments, which leads to the possibility that the quark
and lepton mass matrices may have similar texture patterns.\bigskip

We thank Anjan Joshipura and Ernest Ma for discussions. This work was
supported in part by the U.S. Department of Energy under Grant No.
DE-FG03-94ER40837. \bigskip

\pagebreak

\begin{center}
\textbf{Appendix I}
\end{center}

For the sake of completeness we compile below the neutrino mass matrices for 
$t_{1}=1$ for the cases B and D, although, as discussed earlier, they are in
conflict with the neutrino oscillation data. \newline

\medskip

\rule{350pt}{1pt}

$B_{1},B_{4})%
\begin{pmatrix}
m_{1}+m_{2}+m_{3} & \sqrt{-2(m_{1}+m_{3})(m_{2}+m_{3})} & 0 \\ 
\sqrt{-2(m_{1}+m_{3})(m_{2}+m_{3})} & 0 & m_{3} \\ 
0 & m_{3} & 0%
\end{pmatrix}%
$

(Note: extra texture zero develops since $O(s_{2}^{2})$ are neglected.)

\bigskip \rule{350pt}{1pt}

$B_{2},B_{3})%
\begin{pmatrix}
m_{1}+m_{2}+m_{3} & 0 & \sqrt{-2(m_{1}+m_{3})(m_{2}+m_{3})} \\ 
0 & 0 & m_{3} \\ 
\sqrt{-2(m_{1}+m_{3})(m_{2}+m_{3})} & m_{3} & 0%
\end{pmatrix}%
$

(Note: extra texture zero develops since $O(s_{2}^{2})$ are neglected.)

\bigskip \rule{350pt}{1pt}

$D_{1})%
\begin{pmatrix}
m_{1} & 0 & 0 \\ 
0 & \frac{m_{3}+m_{2}}{2} & \frac{m_{3}-m_{2}}{2} \\ 
0 & \frac{m_{3}-m_{2}}{2} & \frac{m_{3}+m_{2}}{2}%
\end{pmatrix}%
$

\rule{350pt}{1pt}

$D_{2})%
\begin{pmatrix}
m_{1}+m_{2}-m_{3} & 0 & -\sqrt{2}\sqrt{%
m_{3}(m_{1}+m_{2}-2m_{3})-(m_{1}m_{2}-m_{3}^{2})} \\ 
0 & \frac{m_{1}m_{2}-m_{3}^{2}}{m_{1}+m_{2}-2m_{3}} & 0 \\ 
-\sqrt{2}\sqrt{m_{3}(m_{1}+m_{2}-2m_{3})-(m_{1}m_{2}-m_{3}^{2})} & 0 & \frac{%
2m_{3}(m_{1}+m_{2}-2m_{3})-(m_{1}m_{2}-m_{3}^{2})}{m_{1}+m_{2}-2m_{3}}%
\end{pmatrix}%
$

\bigskip \rule{350pt}{1pt}

$D_{3})%
\begin{pmatrix}
m_{1}+m_{2}-m_{3} & \sqrt{2}\sqrt{%
m_{3}(m_{1}+m_{2}-2m_{3})-(m_{1}m_{2}-m_{3}^{2})} & 0 \\ 
\sqrt{2}\sqrt{m_{3}(m_{1}+m_{2}-2m_{3})-(m_{1}m_{2}-m_{3}^{2})} & \frac{%
2m_{3}(m_{1}+m_{2}-2m_{3})-(m_{1}m_{2}-m_{3}^{2})}{m_{1}+m_{2}-2m_{3}} & 0
\\ 
0 & 0 & \frac{m_{1}m_{2}-m_{3}^{2}}{m_{1}+m_{2}-2m_{3}}%
\end{pmatrix}%
$

\bigskip \rule{350pt}{1pt}

$D_{4})%
\begin{pmatrix}
0 & \frac{-2m_{2}m_{1}+m_{3}(m_{1}+m_{2}+m_{3})}{2\sqrt{-2m_{1}m_{2}}} & 
\frac{2m_{2}m_{1}+m_{3}(m_{1}+m_{2}+m_{3})}{2\sqrt{-2m_{1}m_{2}}} \\ 
\frac{-2m_{2}m_{1}+m_{3}(m_{1}+m_{2}+m_{3})}{2\sqrt{-2m_{1}m_{2}}} & 0 & 
\frac{1}{2}(-m_{1}-m_{2}+m_{3}) \\ 
\frac{2m_{2}m_{1}+m_{3}(m_{1}+m_{2}+m_{3})}{2\sqrt{-2m_{1}m_{2}}} & \frac{1}{%
2}(-m_{1}-m_{2}+m_{3}) & m_{1}+m_{2}+m_{3}%
\end{pmatrix}%
$

\bigskip \rule{350pt}{1pt}

$D_{5})%
\begin{pmatrix}
0 & \frac{-2m_{2}m_{1}-m_{3}(m_{1}+m_{2}+m_{3})}{2\sqrt{-2m_{1}m_{2}}} & 
\frac{2m_{2}m_{1}-m_{3}(m_{1}+m_{2}+m_{3})}{2\sqrt{-2m_{1}m_{2}}} \\ 
\frac{-2m_{2}m_{1}-m_{3}(m_{1}+m_{2}+m_{3})}{2\sqrt{-2m_{1}m_{2}}} & 
m_{1}+m_{2}+m_{3} & \frac{1}{2}(-m_{1}-m_{2}+m_{3}) \\ 
\frac{2m_{2}m_{1}-m_{3}(m_{1}+m_{2}+m_{3})}{2\sqrt{-2m_{1}m_{2}}} & \frac{1}{%
2}(-m_{1}-m_{2}+m_{3}) & 0%
\end{pmatrix}%
$

\bigskip \rule{350pt}{1pt}

$D_{6})%
\begin{pmatrix}
0 & \sqrt{\frac{-m_{1}m_{2}}{2}} & -\sqrt{\frac{-m_{1}m_{2}}{2}} \\ 
\sqrt{\frac{-m_{1}m_{2}}{2}} & m_{1}+m_{2} & 0 \\ 
-\sqrt{\frac{-m_{1}m_{2}}{2}} & 0 & m_{1}+m_{2}%
\end{pmatrix}%
$

\bigskip \rule{350pt}{1pt}

\pagebreak

\bigskip $D_{7})%
\begin{pmatrix}
m_{1}+m_{2}-m_{3} & \frac{m_{1}m_{2}-m_{3}^{2}}{\sqrt{2}\sqrt{%
m_{3}(m_{1}+m_{2}-2m_{3})-(m_{1}m_{2}-m_{3}^{2})}} & \frac{%
2m_{3}(m_{1}+m_{2}-2m_{3})-(m_{1}m_{2}-m_{3}^{2})}{\sqrt{2}\sqrt{%
m_{3}(m_{1}+m_{2}-2m_{3})-(m_{1}m_{2}-m_{3}^{2})}} \\ 
\frac{m_{1}m_{2}-m_{3}^{2}}{\sqrt{2}\sqrt{%
m_{3}(m_{1}+m_{2}-2m_{3})-(m_{1}m_{2}-m_{3}^{2})}} & 0 & 0 \\ 
\frac{2m_{3}(m_{1}+m_{2}-2m_{3})-(m_{1}m_{2}-m_{3}^{2})}{\sqrt{2}\sqrt{%
m_{3}(m_{1}+m_{2}-2m_{3})-(m_{1}m_{2}-m_{3}^{2})}} & 0 & 2m_{3}%
\end{pmatrix}%
$

\rule{350pt}{1pt}

\bigskip $D_{8})%
\begin{pmatrix}
m_{1}+m_{2}-m_{3} & \frac{-2m_{3}(m_{1}+m_{2}-2m_{3})+(m_{1}m_{2}-m_{3}^{2})%
}{\sqrt{2}\sqrt{m_{3}(m_{1}+m_{2}-2m_{3})-(m_{1}m_{2}-m_{3}^{2})}} & \frac{%
m_{3}^{2}-m_{1}m_{2}}{\sqrt{2}\sqrt{%
m_{3}(m_{1}+m_{2}-2m_{3})-(m_{1}m_{2}-m_{3}^{2})}} \\ 
\frac{-2m_{3}(m_{1}+m_{2}-2m_{3})+(m_{1}m_{2}-m_{3}^{2})}{\sqrt{2}\sqrt{%
m_{3}(m_{1}+m_{2}-2m_{3})-(m_{1}m_{2}-m_{3}^{2})}} & 0 & 0 \\ 
\frac{m_{3}^{2}-m_{1}m_{2}}{\sqrt{2}\sqrt{%
m_{3}(m_{1}+m_{2}-2m_{3})-(m_{1}m_{2}-m_{3}^{2})}} & 0 & 2m_{3}%
\end{pmatrix}%
$

\newpage

\begin{center}
\bigskip \underline{\textbf{References}}
\end{center}

\begin{enumerate}
\item[{[1]}] See e.g. P. Ramond, R.G. Roberts and G.G. Ross, Nucl. Phys. B406
(1993) 19; B.R. Desai and D.P. Roy, Phys. Rev. D58 (1998) 113007.

\item[{[2]}] S.C. Gibbons, R.N. Mohapatra, S. Nandi and A. Raychaudhuri,
Phys. Lett. B430 (1998) 296; S. Mohanty, D.P. Roy and U. Sarkar, Phys. Lett.
B445 (1998) 185; V. Barger and K. Whisnant hep-ph/0006235, in ``Current
Aspects of Neutrino Physics'', Springer-Verlag, Hamburg (2000).

\item[{[3]}] For reviews see G. Altarelli and F. Feruglio, Phys. Rep. 320
(1999) 295; H. Fritzsch and Z. Xing, Prog. Part. Nucl. Phys. 45, (2000) 1.

\item[{[4]}] B.R. Desai, U. Sarkar and A.R. Vaucher, hep-ph/0007346; H.V.
Klapdor-Klaingrothaus and U. Sarkar, Phys. Lett. B532 (2002) 71; T. Hambye,
hep-ph/0201307.

\item[{[5]}] S.K. Kang and C.S. Kim, Phys. Rev. D63 (2001) 113010 .

\item[{[6]}] P.H. Frampton, S.L. Glashow and D. Marfatia, Phys. Lett. B536
(2002) 79.

\item[{[7]}] Z. Xing, Phys. Lett. B530 (2002) 159; see also M. Frigerio and
A.Yu. Smirnov, hep-ph/0207366 and hep-ph/0202247.

\item[{[8]}] Super-Kamiokande Collaboration: S. Fukuda et al., Phys. Rev.
Lett. 85 (2000) 3999 and references therein.

\item[{[9]}] B.R. Desai, and A.R. Vaucher, Phys. Rev. D63 (2001)13001-1

\item[{[10]}] T.K. Kuo, S.W. Mansour, and G.H. Wu, Phys. Rev. D60
(1999)093004; Phys. Lett. B467 (1999) 116; S.H. Chin, T.K. Kuo, and G.H. Wu,
Phys. Rev. D62 (2000) 053014.

\item[{[11]}] A. Bandyopadhay, S. Choubey, S. Goswami and D.P. Roy, Phys.
Lett. B540 (2002) 14; V. Barger, D. Marfatia, K. Whisnant and B.P. Wood,
Phys. Lett. B537 (2002) 179; J.N. Bahcall, M.C. Gonzalez-Garcia and C.
Pena-Garay, hep-ph/0204314; M. Maltoni, T. Schwetz, M.A. Tortola, J.W.F.
Valle, hep-ph/0207227 and hep-ph/0207157.

\item[{[12]}] M. Apollonio et al., Phys. Lett. B466 (1999) 415; F. Boehm et
al., Phys. Rev. D64 (2001) 112001.

\item[{[13]}] H.V. Klapdor-Kleingrothaus et. al., hep-ph/9910205.

\item[{[14]}] Ernest Ma, D.P. Roy and S. Roy, Phys. Lett. B525 (2002) 101.

\item[{[15]}] A. Kageyama, S. Kaneko, N. Shimoyama and M. Tanimoto, Phys.
Lett. B538 (2002) 96.

\item[{[16]}] L. Baudis et al., Phys. Rev. Lett. 83 (1999) 41.
\end{enumerate}

\end{document}